\documentclass[12pt,a4paper]{article}
\usepackage{graphicx}
\newcounter{N}
\begin{document}
\textwidth=135mm
 \textheight=200mm

\begin{center}
{\bfseries EVOLUTION OF ENERGY DENSITY FLUCTUATIONS IN A+A
COLLISIONS \footnote{{\small Talk given at the The Sixth Workshop
on Particle Correlations and Femtoscopy (WPCF2010), Kiev,
September 14 - 18, 2010.}}} \vskip 5mm M.S. Borysova$^{a, b}$,
Iu.A. Karpenko$^a$,  and  Yu.M. Sinyukov$^{a, c}$ \vskip 5mm
{\small {\it $^a$ Bogolyubov Institute for Theoretical Physics, Kiev, 03680, Ukraine}} \\
{\small {\it $^b$ Kyiv Institute for Nuclear Research, Kiev,  03083, Ukraine}}\\
{\small {\it $^c$ ExtreMe Matter Institute EMMI, GSI Helmholtzzentrum
f\"ur Schwerionenforschung GmbH,  64291 Darmstadt,
Germany}}
\end{center}
\vskip 5mm \centerline{\bf Abstract} Two-particle angular
correlation for charged particles emitted in Au+Au collisions at
the center-of-mass of 200 MeV measured at RHIC energies revealed
novel structures commonly referred to as a near-side ridge. The
ridge phenomenon in relativistic A+A collisions is rooted probably
in the initial conditions of the thermal evolution of the system.
In this study we analyze the evolution of the bumping transverse
structure of the energy density distribution caused by
fluctuations of the initial density distributions that could lead
to the ridge structures. We suppose that at very initial stage of
collisions the typical one-event structure of the initial energy
density profile can be presented as the set of longitudinal tubes,
which are boost–invariant in some space-rapidity region and are
rather thin. These tubes have very high energy density comparing
to smooth background density distribution. The transverse velocity
and energy density profiles at different times of the evolution
till the   chemical freeze-out (at the temperature T=165 MeV) will
be reached by the system
 are calculated for sundry initial scenarios.

PACS number(s): 25.75.Gz, 24.10.Nz
\vskip 10mm
\section{\label{sec:intro}Introduction}
Measurements at the Relativistic Heavy Ion Collider (RHIC,
Brookhaven) have revealed that the long range structure of
two-particle angular correlation functions is significantly
modified by the presence of the hot and dense matter formed in
relativistic heavy ion collisions \cite{Alver}. Novel correlation
structures over large pseudorapidity interval $\Delta y$ were
observed in azimuthal correlations for the intermediate particle
transverse momenta $p_T\approx 1-5$ GeV/c \cite{Alver2, Abelev}.
First the striking "ridge" events were revealed in studies of the
near-side spectrum of correlated pairs of hadrons by the STAR
collaboration \cite{Horner}. The spectrum of correlated pairs on
the near side (defined by the trigger particle direction) extends
across the entire detector acceptance in pseudorapidity interval
of order $\Delta\eta\sim 2$ units and is strongly collimated for
azimuthal angles. Here $\Delta\eta$ is the difference in
pseudo-rapidity  $ \eta = -\ln\left(\tan\left(\theta /
2\right)\right)$, where $\theta$ is the polar angle relative to
the beam axis, between two particles and $\Delta\varphi$ is the
difference in their azimuthal angle $\varphi$ (in radians). The
analyzes of measurements by the PHENIX \cite{Adare} and PHOBOS
\cite{Wosiek} collaborations confirmed the STAR results.

The discovery of the ridge has aggravated quantitative theoretical
analyzes which propose rather different explanations
\cite{Dumitru, Hama, Schenke}. The first, \cite{Dumitru} treats
the ridge as an initial-state effect. The authors point out that
the enhanced two-particle correlations are a natural consequence
of the correlations in the classical color fields responsible for
multi-particle production in relativistic heavy-ion collisions.
Due to fluctuations of color charges in colliding nuclei the
longitudinally boost-invariant and transversally inhomogeneous
structure of the matter can be formed. When it expends
hydrodynamically it could lead to the ridges \cite{Hama}. The
others explore a final-state effect as the origin of the ridge
\cite{Schenke}.The evolution of the system and other later-stage
effects can modify these correlations, which, in fact, are
associated with fluctuations in the energy densities at the final
stage. The recent measurements by the CMS Collaboration \cite{CMS}
observed unexpected effect in  proton-proton collisions at the
LHC. The clear and significant "ridge"-structure emerges at
$\Delta$$\varphi\approx$ 0 extending to $\left|\Delta\eta\right|$
of at least 4 units. This novel feature of the data has never been
seen in two-particle correlation functions in pp or $p\bar p$
collisions before. This unusually elongate structure – the ridge
-remains after removal of elliptic flow and ordinary jet
correlations \cite{Abelev2}. This is faced with the problem of
causality which, probably can be solved only if one supposes that
the ridge phenomenon in relativistic A+A collisions is rooted in
the initial conditions of the thermal evolution of the system. The
aim of this study is to check this hypothesis by an analysis of
the developing energy density in the system which at very initial
stage of collisions has transversally bumping tube-like
fluctuations with boost-invariant homogeneous structure within
some space-rapidity region.

\section{Calculation of energy density distributions}

The numerical results presented in this section were obtained on
the basis of original 3D ideal hydro-code, described in details in
\cite{Akkelin}. The analysis is based on hydrodynamic approach to
A+A collisions and considered within the Boltzmann equations. It
is consistent with conservations laws and accounts for the opacity
effects. The hydrodynamic evolution starts at the time $\tau_0$.
We use Bjorken-type initial conditions at  $\tau_0$:
boost-invariance of the system in longitudinal direction, initial
longitudinal flow $v_L=z/t$  without transverse collective
expansion. In present calculation we compare the
transverse-velocity profile of hydrodynamic flow and energy
density profile which evolve in time till the chemical freeze-out
(T = 165 MeV) in different initial scenarios. The one of them
corresponds to the smooth Gaussian profile with radius R and
energy density as it was considered in \cite{Sinyukov} at
$\tau_0=0.2$ fm/c. The other scenarios are based on transversally
bumping tube-like initial conditions at  $\tau_0$. These tubes are
rather thin transversally and relatively long in the direction of
beam axis; with radii $a_i = 1$ fm. The general energy density
distribution at $\tau_0$  could be written as:

\begin{eqnarray}\label{edd}
  E = E_{b}e^{-\frac{x^2+y^2}{R^2}}+\sum\limits_{i=0}^{N_t}  E_i e^{-\frac{(x-x_i)^2+(y-y_i)^2}{a_i^2}},
\end{eqnarray}
\begin{eqnarray}
R_i=x_i^2=y_i^2
\end{eqnarray}
where $E_b$ is the maximum of average energy density distribution,  $E_i$ are the maxima of tube-like fluctuations, $R_i$  are the positions of the fluctuation locations and $N_t$ is the number of tubes.
Instead of a study of the result over very many fluctuations, that should be finally averaged over azimuthal angular (it brings symmetry), we will be based here on the possible typical, or "representative", initial fluctuation which are already maximally symmetric in azimuthal plane. The following initial configurations are considered:

\begin{list}{\roman{N}}{\usecounter{N}}
\item The configuration without fluctuation: distribution of
initial energy density corresponds to the Gauss distribution with
R = 5.4 fm and maximum energy density at r = 0 is $E_b $ = 90
GeV/fm$^3$; \item The configuration with one tube (fluctuation) in
the center: energy density profile is the Gauss distribution with
a = 1.0 fm and the maxima value 270 GeV/fm$^3$;


\begin{center}
\begin{figure}[!hb]
\mbox{
      \begin{picture}(300,350)(0,0)
      \put(50,0){\includegraphics[width=0.8\textwidth]{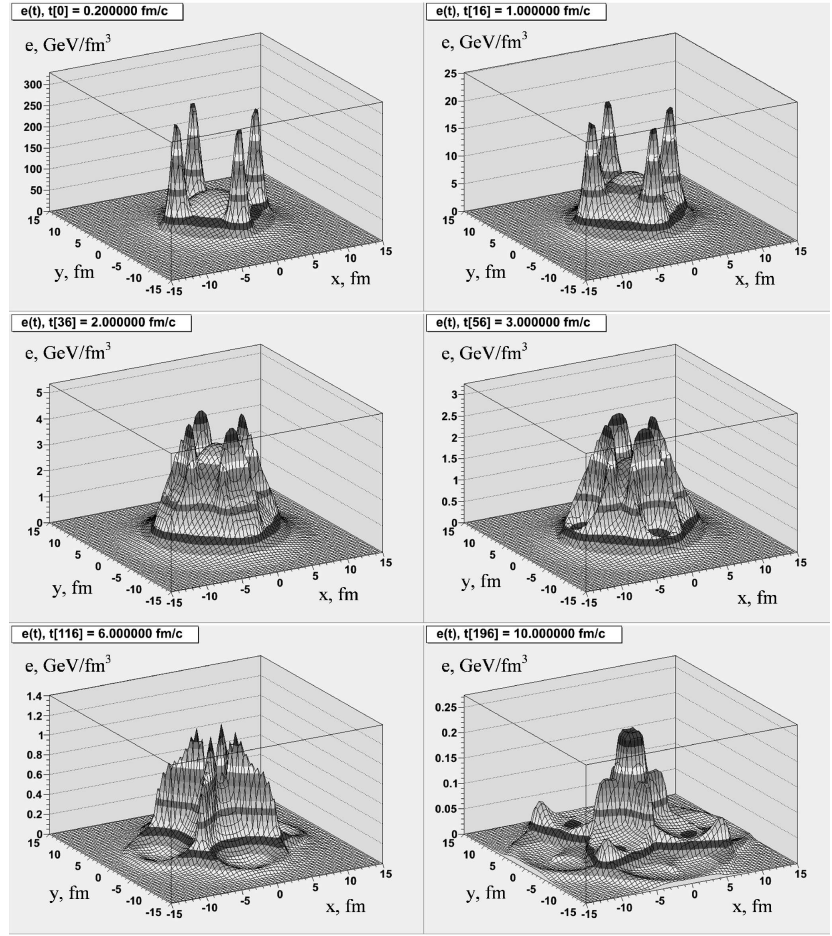}}
      \end{picture}
}
\caption{3D plots of energy density profiles with 4 tube-like
 initial conditions for $\tau$ = 0.2, 1, 2, 3, 6 and 10 fm/c}
\end{figure}
\end{center}

\item  The configuration with one tube (fluctuation) shifted from
the center: $E_b$  = 90 Gev/fm$^3$; R = 5.4 fm; $E_0$ = 270
GeV/fm$^3$; $R_0$ = 3 fm; $a_0$ = 1 fm; The results of
calculations are presented for initial time and for $\tau$ = 1, 2,
10 fm/c ; \item The configuration with four tubes (fluctuations):
$E_b$  = 85 GeV/fm$^3$; R = 5.4 fm; $E_i$= 250 GeV/fm$^3$; $R_i$ =
5.6 fm; $a_i$ = 1 fm; The evolution of energy density profiles is
presented on the Fig. 1; \item The configuration with ten tubes
(fluctuations): $E_b$  = 25 GeV/fm$^3$; R = 5.4 fm; $R_0$ = 0 fm;
$R_i (i\leq 3)$ = 2,8 fm; $R_i (i>3)$  = 4.7 fm; $a_i$ = 1 fm;
$E_i=4E_b\exp(-{R_i^2}/{R^2})$ .
\end{list}

For all the considered cases the traces of the initial
fluctuations - bumping final energy distributions - remain after
the system evolution that should lead to a non-trivial structure
in observed correlations. Besides the energy density profiles the
evolution of transverse velocity profiles of hydrodynamic flow in
different scenarios also was considered. The fluctuations of the
initial conditions result in the fluctuation of the transverse
velocity in the system. At the supposed thermalization  time $\tau
= 1$ fm/c the corresponding fluctuations in the transverse
velocity averaged over azimuthal angular and radius are
approximately 30 \% between the cases when the high density
fluctuation peak is just in the center and when it is shifted
while at the later times ($\tau = 10$ fm/c) it is only 2-3 \% .
This may lead to the important conclusions as for small
fluctuations of the mean transverse momenta of observed particles,
despite big fluctuations of transverse velocities at
thermalization time.
\section{Conclusions}
The basic hydrokinetic code, proposed in \cite{Akkelin} was modified to include the tube-like initial conditions with the aim to study how the initial correlations in the energy density evolve with time. We found that the effect of the fluctuations of the initial conditions does not wash out during the system expansion that, probably, leads to the ridges structures of the correlations, which are caused by these fluctuations. The evolution of the energy density distributions and velocity profiles are calculated in the framework of the 3D hydrodynamics. Possible physically grounded configurations of initial density profiles are proposed. The first spectra calculations were done in the frameworks of the hydrokinetic model (HKM) \cite{Akkelin, Sinyukov2} which allows describing all the stages of the system evolution as well as a formation of the particle momentum at the decoupling stage. The further investigations on this matter and description of correlations could be done in frames of this approach.
\section{Acknowledgments}
 Yu.S. gives thanks to P. Braun-Munzinger for support this study within EMMI/GSI organizations. The researches were carried
  out in part within the scope of the EUREA: European Ultra Relativistic Energies Agreement (European Research Group
  GDRE: Heavy ions at ultrarelativistic energies) and is supported by the State Fund for
Fundamental Researches of Ukraine (Agreements No. F33/461-2009,
2011)  and National Academy of Sciences of Ukraine (Agreements No
F06-2010, 2011). The numerical calculations were done by using the
cluster and GRID environment of the Bogolyubov Institute for
Theoretical Physics of National Academy of Sciences of Ukraine.

\end{document}